\documentclass[conference]{IEEEtran}
\IEEEoverridecommandlockouts
\usepackage{cite}
\usepackage{amsmath,amssymb,amsfonts}
\usepackage{graphicx}
\usepackage{textcomp}
\usepackage{xcolor}

\usepackage{comment}

\usepackage{algorithm}
\usepackage{algpseudocode}

\usepackage{adjustbox}

\usepackage{subcaption}

\usepackage{enumitem}

\def\BibTeX{{\rm B\kern-.05em{\sc i\kern-.025em b}\kern-.08em
    T\kern-.1667em\lower.7ex\hbox{E}\kern-.125emX}}

\begin{document}

\title{QEdgeProxy: QoS-Aware Load Balancing for IoT Services in the Computing Continuum
}


\author{
    \IEEEauthorblockN{
    Ivan Čilić\IEEEauthorrefmark{1},
    Valentin Jukanović\IEEEauthorrefmark{1},  
    Ivana Podnar Žarko\IEEEauthorrefmark{1}, 
    Pantelis Frangoudis\IEEEauthorrefmark{2}}
    Schahram Dustdar\IEEEauthorrefmark{2}\\
    \IEEEauthorblockA{\IEEEauthorrefmark{1}Faculty of Electrical Engineering and Computing, University of Zagreb, Zagreb, Croatia
    \\\{ivan.cilic, valentin.jukanovic, ivana.podnar\}@fer.hr} \\
    \IEEEauthorblockA{\IEEEauthorrefmark{2}Distributed Systems Group, TU Wien, Vienna, Austria
    \\\{p.frangoudis, dustdar\}@dsg.tuwien.ac.at}
}

\maketitle

\begin{abstract}
While various service orchestration aspects within Computing Continuum (CC) systems have been extensively addressed, including service placement, replication, and scheduling, an open challenge lies in ensuring uninterrupted data delivery from IoT devices to running service instances in this dynamic environment, while adhering to specific Quality of Service (QoS) requirements and balancing the load on service instances. To address this challenge, we introduce \textit{QEdgeProxy}, an adaptive and QoS-aware load balancing framework specifically designed for routing client requests to appropriate IoT service instances in the CC. QEdgeProxy integrates naturally within Kubernetes, adapts to changes in dynamic environments, and manages to seamlessly deliver data to IoT service instances while consistently meeting QoS requirements and effectively distributing load across them. This is verified by extensive experiments over a realistic K3s cluster with instance failures and network variability, where QEdgeProxy outperforms both Kubernetes built-in mechanisms and a state-of-the-art solution, while introducing minimal computational overhead.
\end{abstract}


\section{Introduction} 
\label{intro}

Computing Continuum (CC) is a highly dynamic edge-to-cloud environment characterized by continuous changes in the states and locations of IoT devices and edge nodes, where diverse services running on the edge nodes have to be scheduled, deployed and managed to meet potentially diverse Quality of Service (QoS) objectives, such as high availability or low response times, or service-specific objectives such as inference accuracy of ML models. 
A significant body of work~\cite{DBLP:conf/infocom/WojciechowskiOL21, DBLP:conf/netsoft/0001WVT19, DBLP:journals/tiot/ToczeFPN23, Sensors:Krivic, DBLP:conf/ucc/PusztaiNMCRDVXZ22} on QoS-aware service orchestration in CC focuses on optimal service placement on compute nodes, often considering the volatility of the environment typically experienced at the edge.
Once services are deployed across the CC, the challenge arises on how to ensure continuous data delivery from IoT devices to the corresponding distributed service instances for further processing, e.g. data filtering, aggregation or inference, while taking into account the dynamic nature of the CC --- potential node and service failures, service migration, as well as service network QoS degradation. 

In practice, each service is associated to a set of QoS requirements specifying Service Level Objectives (SLOs) that must be met for all clients using the service~\cite{DBLP:conf/ucc/PusztaiNMCRDVXZ22}. These requirements are based on various parameters, including latency, throughput, and security. In this paper, we focus on enabling \textit{adaptive and QoS-aware IoT data routing in the CC} without incurring significant overhead to resource-constrained IoT devices (i.e., clients) and services deployed in the continuum, and to enable \textit{seamless integration within existing orchestration platforms}, such as Kubernetes~\cite{kubernetes}. Hence, we introduce \emph{QEdgeProxy}, a QoS-aware load balancer designed as a QoS agent for IoT clients. Its primary function is to identify service instances across the continuum that could fulfill the QoS requirements for a given service, forming a ``QoS pool'', and efficiently distribute requests among them. 

What sets our approach apart from related work, such as~\cite{7912240, DBLP:conf/ccgrid/FahsP19, s22082869}, is our intentional design of the proxy to distribute the workload evenly across instances capable of meeting the QoS requirements specified for a particular service, rather than identifying instances that optimize QoS or perform trade-offs between QoS and load balancing, which can lead to instance overload or QoS degradation. With this technique, we minimize the risk of overloading service instances while ensuring that clients continuously receive satisfactory QoS. Note that our solution is tailored to integrate seamlessly with the complex Kubernetes environment which offers many services and built-in mechanisms that need to be taken into account when designing solutions for QoS-aware service orchestration tailored to the CC.   
Our contributions can be summarized as follows: 
\begin{itemize}[leftmargin=8pt]
  \item We present the concept, architecture and mechanisms of \textit{QEdgeProxy}~(\S\,\ref{proxy}).
  \item We present its implementation, that naturally fits within the Kubernetes environment~(\S\,\ref{impl}). QEdgeProxy is implemented as an external component running on every node and collecting cluster information to optimize routing decisions.
  \item We evaluate QEdgeProxy in an emulated network environment on top of a K3s~\cite{k3s} cluster through a real-world IoT use case~(\S\,\ref{evaluation}). Evaluation results show that QEdgeProxy continuously satisfies QoS requirements while performing load balancing on the target instances. Compared to the Kubernetes built-in mechanisms and state-of-the-art solutions, it provides better load balancing while maintaining a high rate of requests that meet the QoS requirements.
\end{itemize}
\par
\section{Related work} 
\label{related_work}

Existing relevant work primarily focuses on QoS-aware service scheduling \cite{DBLP:conf/infocom/WojciechowskiOL21, DBLP:conf/netsoft/0001WVT19, DBLP:journals/tiot/ToczeFPN23,DBLP:conf/ucc/PusztaiNMCRDVXZ22, Sensors:Krivic}, but rarely on QoS-aware continuous delivery of IoT data or requests to deployed service instances. For the latter, Kapsalis et al.~\cite{7912240} propose a fog request broker that implements target instance scoring based on the current utilization, latency, and battery state of each target host (applicable to mobile devices). Similarly, Fahs and Pierre~\cite{DBLP:conf/ccgrid/FahsP19} introduce \textit{proxy-mity}, a proxy deployed on each Kubernetes node to intercept client requests and route them to the target service instance based on network latency. The routing decision is based on an adjustable parameter that represents the desired trade-off between pure load-balancing and pure proximity-based routing. Nguyen et al.~\cite{s22082869} present Resource Adaptive Proxy (RAP), which performs regular checks on the resource status of individual pods\footnote{A Kubernetes Pod is the smallest deployable unit in Kubernetes, representing a group of one or more containers that share the same network namespace and storage, and are scheduled together on the same node.} and the network status between worker nodes to make load-balancing decisions, prioritizing local request handling. RAP and proxy-mity require replacing the default Kubernetes functionality (\textit{kube-proxy}). Other works~\cite{Rejiba21,Karagiannis23} use Reinforcement Learning for service instance selection, but do not explicitly express and enforce QoS constraints.

Our approach exhibits two crucial distinctions. 
\emph{First}, we explicitly address the diverse QoS requirements of different services. For instance, some applications, such as those supporting Connected and Automated Mobility, target latencies from below 10\,ms~\cite{Boban18} to a few hundred milliseconds~\cite{Alam23}, depending on the scenario, while others require less stringent latency, e.g., below 1000\,ms. QEdgeProxy, equipped with knowledge of these QoS requirements, constructs a \textit{per-service QoS pool} comprising instances that fulfill these specifications, enabling dynamic load balancing across these nodes. This strategy ensures continuous QoS adherence while minimizing the risk of overloading service instances, unlike alternative solutions that rigidly distribute requests within the same set of nodes in the proximity of the source node to optimize QoS, or occasionally randomly select nodes, potentially leading to QoS violations.
\emph{Second}, QEdgeProxy promptly reacts to QoS degradation, also differentiating itself from solutions~\cite{7912240, DBLP:conf/ccgrid/FahsP19, s22082869} that use QoS approximations predominantly based on network latency, thus not capturing processing delays or the waiting time induced by the target instance's queue. By taking into account the actual measurements of QoS parameters, our approach ensures more effective routing decisions, as shown by our experiments.
\section{QEdgeProxy} 
\vspace{-0.8mm}
\label{proxy}
\subsection{Architecture and mechanisms}
To ensure continuous data delivery to IoT services while preserving the targeted QoS for the IoT clients, we propose \textit{QEdgeProxy}, a distributed QoS-aware load balancer tailored to the CC. Its primary functions include (i) dynamically maintaining a set of service instances that meet the targeted QoS for a given service, and (ii) forwarding service requests to these instances while performing load balancing. QEdgeProxy serves as a QoS agent for IoT clients, and acts as an external routing component, i.e., an intermediary between clients and IoT services across the CC. This design relieves clients of the responsibility to discover and choose service instances to connect with. What sets QEdgeProxy apart from other edge-aware proxies is that it focuses primarily on \textit{adaptively meeting QoS requirements for its clients}, rather than solely selecting instances offering the best QoS at a particular time. This approach enables the integration of load balancing techniques with request forwarding to mitigate the risk of overloading the processing nodes while adhering to QoS constraints. As QoS largely depends on the processing node where the instance is deployed, and considering that different services can have different QoS requirements, this approach broadens the range of nodes that can be selected for processing a request, potentially leading to enhanced load balancing efficiency.  

\begin{figure}[htbp]
\centering
\includegraphics[width=0.5\columnwidth]{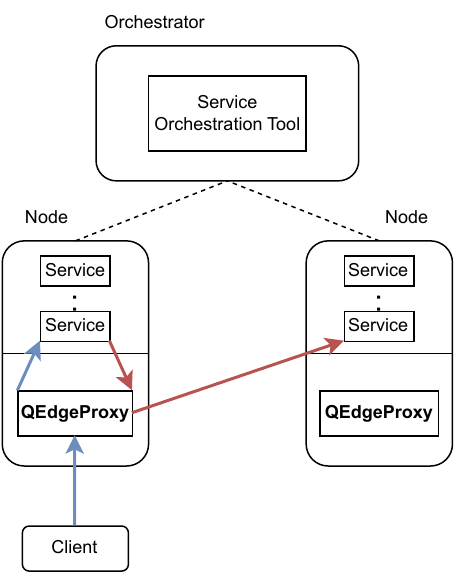}
\caption{QEdgeProxy within the CC.}
\label{fig:edge-proxy}
\vspace{-0.2cm}
\end{figure}

As shown in \figurename~\ref{fig:edge-proxy}, QEdgeProxy is a component deployed on every node within the CC. It integrates seamlessly into existing edge solutions, which are typically managed by a central \textit{orchestrator} component, as in existing service orchestration tools like Kubernetes~\cite{kubernetes}. Services are deployed and executed on heterogeneous edge nodes, distributed across different levels of the CC hierarchy. These services need to be autonomous and portable to facilitate service migrations within this dynamic environment. Therefore, services are commonly deployed and run within Virtual Machines (VMs) or containers, such as Docker-based ones. 

QEdgeProxy is designed to be exposed on each node in the CC and to run just as any other service using the service orchestration tool. A client initiating requests to an IoT service, e.g., a data filtering, data aggregation or ML inference service, must connect to its dedicated QEdgeProxy to be able to access that service. This proxy then forwards the request to the appropriate service instance. As shown in \figurename~\ref{fig:edge-proxy}, the proxy enables both client-to-service communication (blue arrows) and service-to-service communication (red arrows). A service running within the CC, e.g., a data filtering service, can forward the resulting data stream to services running on other nodes, e.g. to an ML inference service. In case of such service-to-service data routing, a request is forwarded through the QEdgeProxy instance running on the same node as the initiating service. External clients, primarily IoT devices operating outside an edge cluster, connect to the QEdgeProxy running on a node in their proximity, i.e. their dedicated QEdgeProxy. A potential solution involves connecting to the proxy running on the node with the shortest network distance from the client, but more sophisticated mechanisms are beyond the scope of this work. 

The overall functionality of QEdgeProxy can be summarized as the process of constructing and maintaining a \textit{QoS pool} for a given service, i.e. a set of service instances capable of fulfilling the specified QoS requirements. This pool serves as the basis for routing incoming requests to the most suitable instances, ensuring uninterrupted data delivery and adherence to QoS constraints. The functionality of QEdgeProxy is thus defined by two main algorithms: the routing algorithm and the monitoring algorithm.

\begin{algorithm}[htbp]
\caption{Request routing workflow.}
\label{alg:routing}
\begin{algorithmic} [1]
{\ttfamily \scriptsize{
\Procedure{RequestRouting}{$request,service$}
\State $serviceInstance \gets getInstanceFromQosPool(service)$
\State $qos \gets sendRequest(request, serviceInstance)$
\State $storeQosMeasurements(serviceInstance,qos)$
\EndProcedure
}}
\end{algorithmic}
\end{algorithm}

Algorithm~\ref{alg:routing} defines a high-level request routing logic executed by a QEdgeProxy instance. When a request arrives for a target service, the QEdgeProxy selects a service instance from the previously built QoS pool and forwards the request to the selected instance. After forwarding the request, the proxy measures the QoS attained and stores the observed values so that this information can be used for maintaining the QoS pool. 

To keep the QoS pool up-to-date, the QEdgeProxy performs event-based system monitoring. There are four events that can trigger actions upon which QoS pools are updated:
\begin{enumerate}[leftmargin=14pt]
    \item \textbf{Initial QoS Pool Creation:} Upon receiving the first request for a given service, the proxy establishes a new dedicated QoS pool based on initial estimates, such as environmental conditions or historical interactions with instances of similar services running on the same nodes.
    \item \textbf{Instance State Updates:} Through the connection with the orchestrator, the proxy continuously monitors the state of service instances, including instances joining or leaving the network. Upon detecting any instance state changes, the proxy updates the relevant QoS pool accordingly. 
    \item \textbf{QoS Measurement Updates:} After forwarding a request to a service instance and measuring its QoS, the proxy incorporates these measurements into the corresponding QoS pool, ensuring that the pool reflects the actual performance of the instances.
    \item \textbf{Environment Monitoring and Reactive Updates:} QEdgeProxy continuously monitors the environment for any significant changes, such as network disruptions or resource fluctuations on CC nodes. Upon detecting such changes, the proxy triggers a comprehensive evaluation of all QoS pools affected by the environmental alterations and updates them accordingly to maintain QoS consistency.
\end{enumerate}

\subsection{Definition of a QoS pool}

Each service is associated with a set of minimum QoS requirements that define the acceptable performance levels for the service to meet its intended purpose for any client. These requirements commonly encompass various aspects of service performance, including latency, throughput, reliability, availability, security, and others.

Each service consists of multiple replicas, referred to as instances. Each QEdgeProxy predicts the QoS of a specific instance based on approximations and its past behavior, e.g., measured network latency based on past interactions with an instance. Various techniques, which are outside the scope of this work, can be utilized for predicting QoS based on recent QoS measurements and interactions with the node running the instance, e.g. LSTMs or other machine learning techniques. Finally, the \textbf{QoS pool} is a subset of service instances that are likely to satisfy the QoS requirements.
\section{QEdgeProxy within Kubernetes: implementation considerations} 
\label{impl}

We present the QEdgeProxy implementation within the Kubernetes ecosystem. Kubernetes, a widely adopted, powerful, and stable container orchestration platform, provides a robust foundation for our solution. Additionally, its lightweight distributions, such as K3s~\cite{k3s}, are particularly well-suited for edge computing environments, as demonstrated by studies such as~\cite{cilicPerf}.

Several built-in and open-source mechanisms in Kubernetes can be used to implement adaptive and custom routing at the edge. One of them is the NodePort service type which exposes a dedicated port on each node in the cluster to access a particular service: thus, an end device can connect to any node in the CC to send a request to this service. However, behind the NodePort service, kube-proxy equally balances requests to the target pods, without any possibility to implement custom parameters, such as selecting nearby pods. Another Kubernetes feature, the NetworkPolicy resource~\cite{network-policy}, enables defining rules for communication between pods. For instance, setting \textit{externalTrafficPolicy: Local} specifies that external traffic should be processed locally on the node that receives it. However, this solution only allows processing a request on the node which received it, and if no adequate pod is found, the request is discarded. On the other hand, Istio \cite{istio} is an open service management platform that implements the service mesh architectural pattern, simplifying cluster traffic and security management. It works by injecting a specific container into pods, known as a "sidecar proxy," which handles communication with other services and allows for defining additional traffic routing rules, security measures, and visibility. While Istio supports routing settings considering the locality of individual pods, it differs from adaptive routing, as it requires manual setting of regions and zones for resources. This approach contradicts the principle of adaptive routing, making Istio unsuitable for implementing the QEdgeProxy.

\begin{figure}[htbp]
\centering
\includegraphics[width=3in]{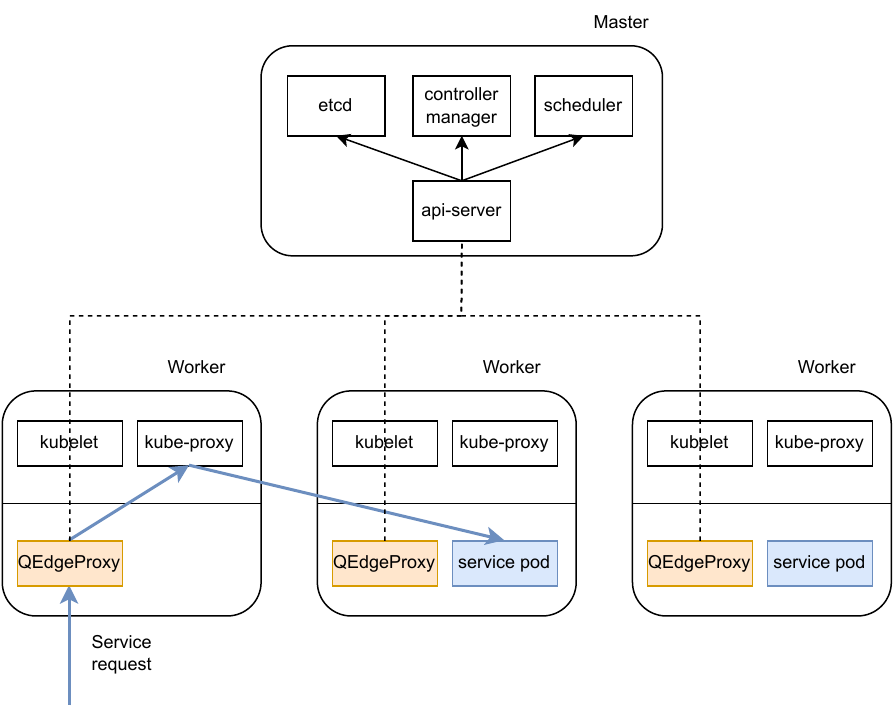}
\caption{QEdgeProxy within a Kubernetes environment.}
\label{fig:ep-k8s}
\vspace{-2mm}
\end{figure}

Based on the deficiencies of existing solutions, we have decided to implement a custom external component for adaptive request routing and deploy it in Kubernetes as a DaemonSet. DaemonSet is a Kubernetes resource which ensures that all worker nodes run a copy of a given pod, which in our case is the QEdgeProxy. 
QEdgeProxy is deployed just like any other pod within a Kubernetes environment, as shown in \figurename~\ref{fig:ep-k8s}. It is implemented as an HTTP server written in Golang\footnote{Due to the double-blind review, the GitHub repository will be added if the paper is accepted.} and it connects to the Kubernetes API server to obtain information about the service instances available in the CC, as well as nodes that are running them. QEdgeProxy continuously observes the state of the CC by subscribing to the events in the Kubernetes API. Clients access a service by sending their HTTP requests to a QEdgeProxy with the service name included in the request header. When an instance is selected by the QEdgeProxy, the request is forwarded directly to the pod of the service instance through kube-proxy.

\section{Experimental evaluation} 
\label{evaluation}

To evaluate the effectiveness of our QEdgeProxy implementation, we conducted a comparative experimental evaluation using three different configurations on a K3s cluster, a lightweight distribution of Kubernetes. These configurations differ in the request routing mechanism employed, while maintaining a consistent execution flow. To align with the state of the art and compare our solution against relevant approaches, we focused on exploring latency as a single QoS requirement in this evaluation, although our QEdgeProxy can support more complex QoS specifications. The configurations used in experiments are outlined below: 
\begin{itemize}[leftmargin=8pt]
    \item \textbf{Kubernetes NodePort Service}: This configuration uses the built-in Kubernetes NodePort Service which exposes a port on every node, enabling direct access to IoT services (as described in Section~\ref{impl}). This comparison provides insights into the QEdgeProxy's performance relative to a standard Kubernetes routing solution.
    \item \textbf{Proximity-based routing}: This configuration implements the proximity-based routing algorithm \textit{proxy-mity}, proposed by Fahs and Pierre~\cite{DBLP:conf/ccgrid/FahsP19}, and outlined in Section~\ref{related_work}. This algorithm incorporates a key parameter \(\alpha\) that controls the trade-off between load balancing (when \(\alpha=0.0\)) and pure proximity-based routing (when \(\alpha=1.0\)). Two sub-configurations were evaluated: (i) \textit{proxy-mity 1.0} with \(\alpha=1.0\), prioritizing proximity for minimal latency; and (ii) \textit{proxy-mity 0.8} with \(\alpha=0.8\), achieving a balance between proximity and load balancing. The reason for choosing these values in two sub-configurations is that we wanted to compare our solution first with the configuration that favors proximity and offers the lowest latency, and second with another configuration which favors proximity with a slight trade-off with load balancing, which we believe is the most similar to the QEdgeProxy behavior.
    \item \textbf{QEdgeProxy}: This configuration employs our QEdgeProxy as the primary request routing mechanism from IoT clients to IoT services. As described in Section~\ref{impl}, a QEdgeProxy instance is deployed as a Kubernetes DaemonSet and runs on every node in the cluster, enabling dynamic routing and load balancing to meet the predefined service QoS requirements.
\end{itemize}

\subsection{Experimental setup}

The K3s cluster in which the experiments are conducted consists of 7 nodes, six worker nodes, and one master node. All nodes are implemented as virtual machines equipped with 2 processor cores and 4 GB of RAM. These nodes are interconnected using the Imunes tool~\cite{imunes},  a network emulator/simulator capable of replicating realistic network topologies. We employed Imunes to simulate network delays and instance failures in our experiments.

\begin{figure}[htbp]
\centering
\includegraphics[width=3in]{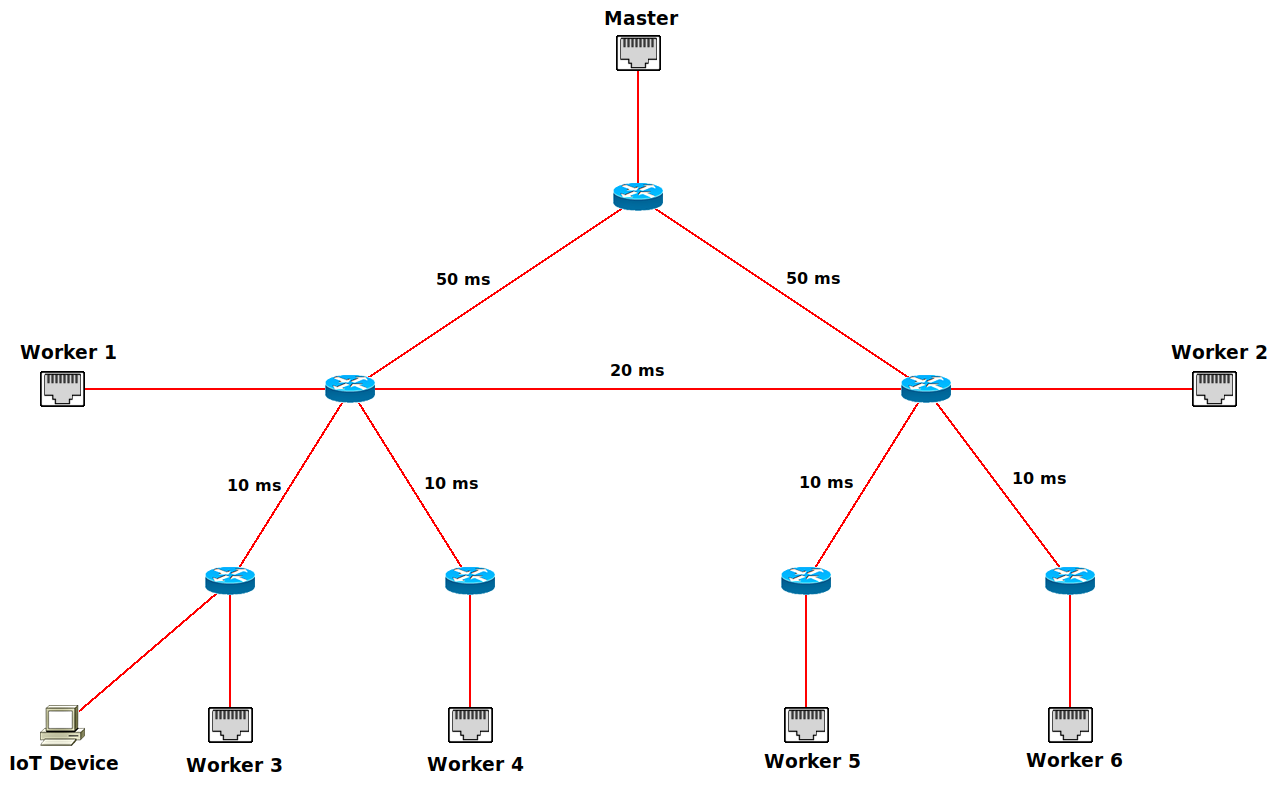}
\caption{Network topology of the evaluation in the Imunes emulator.}
\label{fig:topology}
\vspace{-2mm}
\end{figure}

\figurename~\ref{fig:topology} depicts the network topology used for the experiments, illustrating the assumed network delays of 10, 20, and 50\,ms introduced on the links between routers. These delays mimic the links between the far edge (e.g., LAN), near edge (e.g., a telco-operated Multi-access Edge Computing host) and the cloud. These numbers are roughly based on our round-trip time measurements in (i) a WLAN environment, (ii) over a 5G mobile network connection, and (iii) towards a data center of a global cloud provider in the same continent. The K3s master node is positioned in the cloud, while nodes \textit{Worker 1} and \textit{Worker 2} are placed at the near edge, closer to the cloud. The remaining four worker nodes and the IoT device are deployed at the far edge layer of the CC topology. An IoT device is simulated as an Imunes computer and is connected to the same local network as node \textit{Worker 3}.

\subsection{Experimental evaluation scenarios}

For the purpose of this evaluation, we generated \(\approx 4000\) HTTP POST requests from the IoT device to the proxy running on node \textit{Worker 3}. Each request contains a sensor reading for air humidity, temperature, or CO2 concentration, and thus the HTTP request generator acts as a virtual IoT sensor. All IoT device requests are sent to the service responsible for data validation and processing (Data Processing Service, DPS). We defined a single QoS requirement for DPS: its response time has to be below \textbf{80 milliseconds}. This service notifies the IoT device with a status code about the (un)successful processing of each request, and forwards the processed data to the next service in the data pipeline for advanced analytics, such as an inference service within an ML pipeline for accident predictions. The DPS service instances were deployed within Kubernetes pods on the K3s cluster.

We define DPS to consist of seven instances \(S=\{s_1, \dots, s_7\}\), where \(s_i\) defines that a service instance is running on node \textit{Worker} \emph{i}, except instance \(s_7\) which is deployed on the Master node. Since evaluation scenarios assume that the data are sent from one IoT device and there is no load coming from other parts of the system, the QEdgeProxy uses a simplified load balancing function which assigns the same weights to all instances in the QoS pool. The initial QoS pool is populated with instances choosen based on the QoS approximations derived from network latency. This pool is continuously updated and maintained using recent latency measurements piggybacked on client requests. Two scenarios were devised to assess the performance of the previously described configurations: static and dynamic scenario. 

\subsubsection{Static scenario}
In the static scenario, seven DPS instances were deployed, one per each node. All requests were sent to the proxy running on node \textit{Worker 3} and during the whole experiment, there were no changes in the network nor in instances deployed in the K3s cluster. This scenario evaluates the ability of each configuration to meet the specified QoS requirement under stable conditions.

\subsubsection{Dynamic scenario}
\label{dyn-scenario}
In the dynamic scenario, the state of the pods representing DPS instances or the state of the network is changed after every 750 requests. These changes were designed to simulate real-world events that can impact the QoS of IoT services. The sequence of changes implemented in the dynamic scenario is as follows:
\begin{itemize}[leftmargin=8pt]
    \item \textbf{T0}: At the start of the simulation, the total of 6 DPS instances are deployed on all nodes except node \textit{Worker\,3}.
    \item \textbf{T1}: Instance \(s_3\) is deployed on node \textit{Worker 3}, increasing the total number of DPS instances to seven. 
    \item \textbf{T2}: Instance \(s_3\) becomes overloaded, causing its processing delay to increase by 100\,ms. 
    \item \textbf{T3}: Instance \(s_3\) fails and is removed from the cluster, reducing the number of instances to 6. 
    \item \textbf{T4}: Network latency towards instance \(s_1\) (on the link between \textit{Worker 1} and its router) is increased by 100\,ms. 
\end{itemize}

The purpose of this dynamic scenario is to assess the adaptiveness of the three configurations in handling various events in the CC that can impact QoS: (i) service instance joining and leaving the CC; (ii) changes in processing latency when a node becomes overloaded; and (iii) changes in network conditions, such as increased latency due to network instability.

\subsection{Results}

\subsubsection{Service response time}

The first evaluation parameter is the service response time, as the latency threshold was set as the only QoS requirement for DPS. Table~\ref{table:comparison} compares all configurations by showing average response times and rates of requests that meet the QoS. 

In the static scenario, we can see that the Kubernetes built-in solution NodePort shows by far the worst performance: The largest number of requests handled by NodePort are outside the acceptable QoS range, primarily due to its round-robin load distribution mechanism. This results in a large number of requests being served by instances located far away from the IoT device in terms of network distance, leading to unacceptably high latency. In contrast, both QEdgeProxy and proxy-mity 1.0 consistently maintain low response times which meet the QoS threshold, while proxy-mity 0.8 occasionally fails to meet the required QoS limit. This occasional deviation can be attributed to proxy-mity's trade-off between proximity-based routing and load balancing. To ensure balanced load distribution, proxy-mity may occasionally route requests to further-off instances. Comparing QEdgeProxy to proxy-mity 1.0, we observe that the average response time of proxy-mity 1.0 is significantly lower compared to QEdgeProxy due to the fact that it favors the closest node, while our solution prioritizes meeting the QoS requirements while also performing load balancing. This results in some requests experiencing higher latency, but still within the acceptable range. However, the number of requests meeting the QoS target, which is the most important parameter for the service client, is nearly identical, favoring proxy-mity by a mere 0.11\%. QEdgeProxy thus achieves nearly the same QoS guarantees as the purely latency-centric proxy-mity 1.0, while simultaneously performing the load-balancing of requests which reduces the risk of overloading low-latency nodes.

In the dynamic scenario, NodePort's static load distribution across all available instances again leads to poor performance. Both proxy-mity configurations show more requests exceeding the 80\,ms latency threshold compared to the static scenario. However, QEdgeProxy maintains a high proportion of requests within the threshold, outperforming other solutions. Further explanation is provided in Section~\ref{adapt}.

\begin{table}[ht!]
\centering
\caption{Response time and QoS success comparison: static and dynamic scenario.}
\label{table:comparison}
\begin{tabular}{|l || *{4}{p{1cm}|}} 
 \hline
 \textbf{Configuration} & \multicolumn{2}{p{2cm}|}{\textbf{Average response time}} & \multicolumn{2}{p{2cm}|}{\textbf{Successful request rate (QoS)}} \\ [0.5ex] 
 \hline
 & \textit{Static} & \textit{Dynamic} & \textit{Static} & \textit{Dynamic} \\ 
 \hline
  NodePort & 120,11\,ms & 120,11\,ms  & 28,70\% & 28,70\% \\ 
 \hline
 proxy-mity 0.8 & 20,29\,ms & 61,05\,ms & 91,01\% & 70,78\% \\
 \hline
 proxy-mity 1.0 & \textbf{6,02\,ms} & 51.40\,ms & \textbf{99,97\%} & 76,27\% \\
 \hline
 QEdgeProxy & 26,94\,ms & \textbf{44,04\,ms} & 99,86\% & \textbf{98,86\%} \\ 
 \hline
\end{tabular}
\end{table}

\subsubsection{Load balancing}

To evaluate the effectiveness of load balancing capabilities of each configuration, we compare their request distribution across service instances in the static scenario. This comparison is limited to the static setting since the dynamic scenario introduces fluctuating conditions that may distort the load balancing assessment. Instead, we evaluate load balancing in the context of the dynamic scenario in Section~\ref{adapt}.

\begin{figure}[htbp]
\centering
\includegraphics[width=3in]{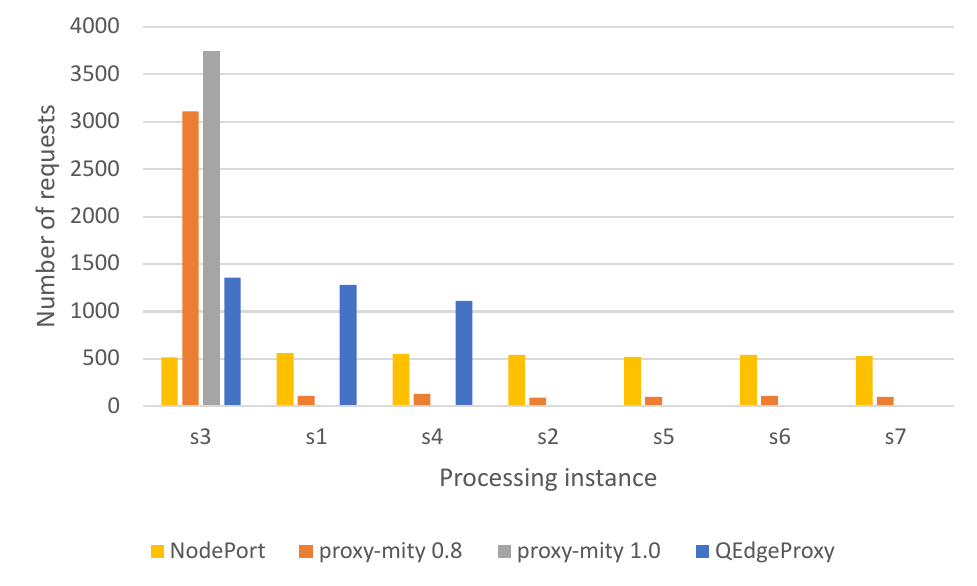}
\caption{Processing instance distribution: static scenario.}
\label{fig:stat-inst}
\end{figure}

\figurename~\ref{fig:stat-inst} shows the distribution of processing instances that served the requests. The instances are sorted based on the network distance from the IoT device to the node running the instance. NodePort evenly distributes requests across all available instances, adhering to the standard Kubernetes round-robin behavior. Proxy-mity 1.0, on the other hand,  sends all requests to the closest instance (\(s_3\)), reflecting its prioritization of proximity. Proxy-mity 0.8 exhibits a mixed strategy, with a majority of requests being routed to instance \(s_3\), while occasionally routing requests to other instances, with an almost equal distribution among the remaining instances. In contrast, QEdgeProxy maintains an even load distribution among all three instances within its QoS pool, ensuring continuous adherence to the desired QoS for IoT clients. 

\subsubsection{Adaptiveness}
\label{adapt}
We finally evaluate the ability of the candidate routing solutions to adapt to changes in the ECC, such as instance failures, changes in processing load, or changes in the underlying network. We evaluate this through the dynamic scenario. 

\begin{figure}[t!]
    \centering
    \begin{subfigure}[t]{0.25\textwidth}
        \centering
        \includegraphics[height=1.1in, width=\textwidth]{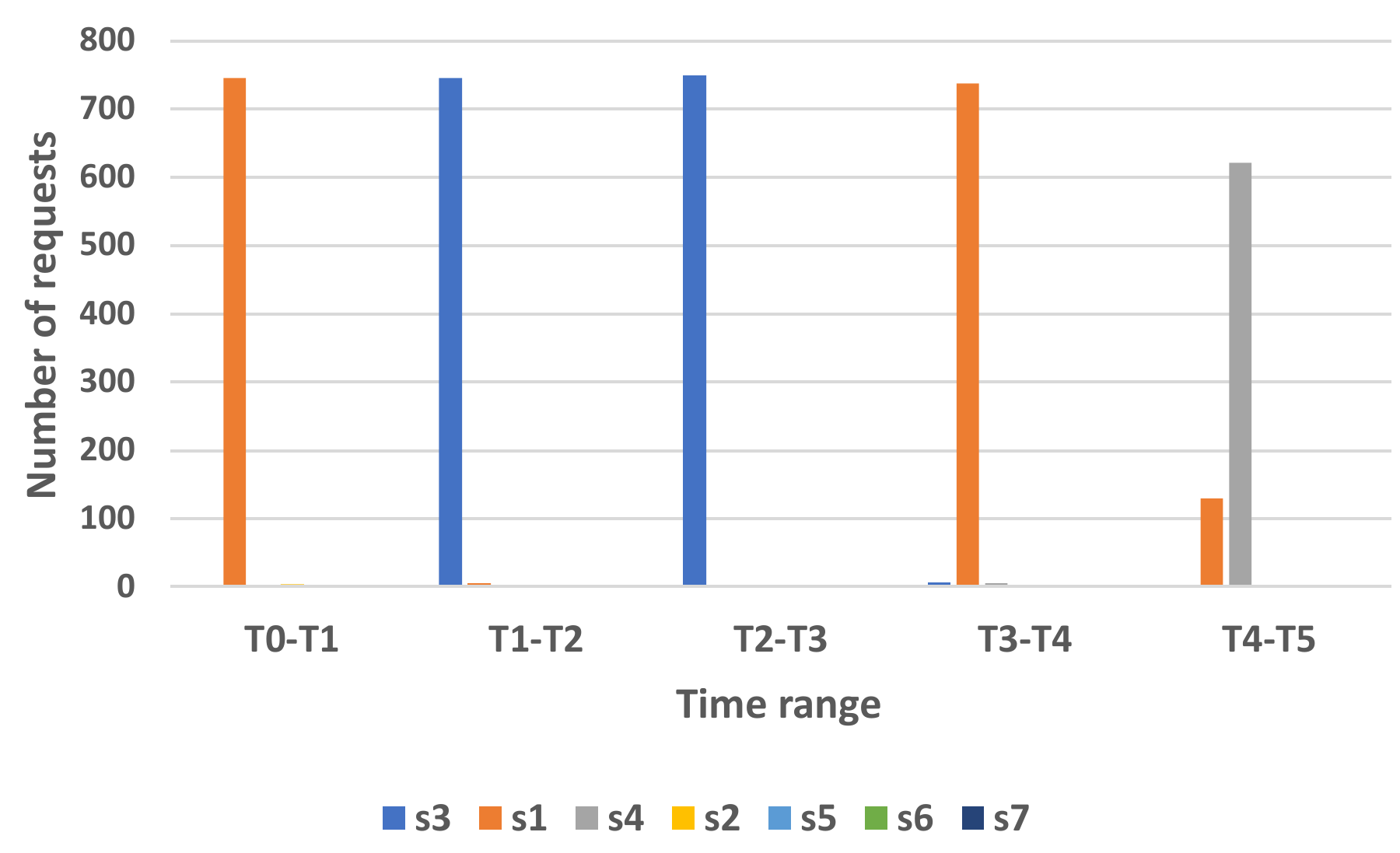}
        \caption{proxy-mity 1.0}
        \label{fig:dyn-node-p10}
    \end{subfigure}%
    ~ 
    \begin{subfigure}[t]{0.25\textwidth}
        \centering
        \includegraphics[height=1.1in, width=\textwidth]{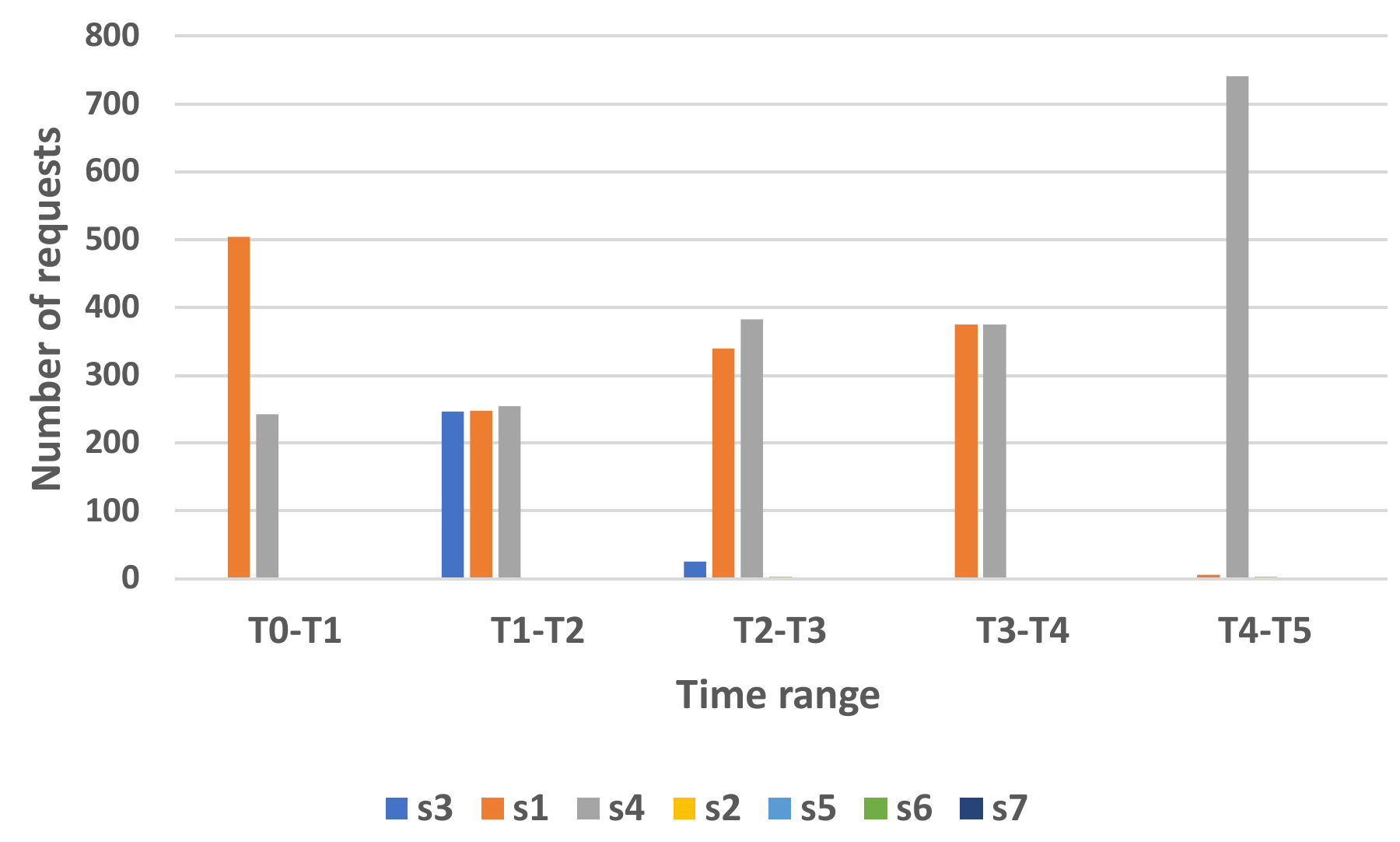}
        \caption{QEdgeProxy}
        \label{fig:dyn-node-ep}
    \end{subfigure}
    \caption{Processing instance distribution per time range: dynamic scenario.}
    \label{fig:dyn-node}
    \vspace{-3mm}
\end{figure}

The graphs in~\figurename~\ref{fig:dyn-node} show how the distribution of requests to processing instances changed over time in the dynamic scenario. The graphs only compare proxy-mity 1.0 and QEdgeProxy configurations as they have the highest QoS success rate (Table \ref{table:comparison}). The time range is set between the five timestamps defined in Section \ref{dyn-scenario}.
The distribution graph of proxy-mity 1.0 (\figurename~\ref{fig:dyn-node-p10}) shows that proxy-mity always favors routing to the closest node based on network latency. Therefore, it did not detect the increase in processing latency between timestamps T2 and T3, and continued to send most of the requests to instance \(s_3\), which resulted in high response times and lowered the number of successful requests in terms of meeting the QoS threshold. On the other hand, QEdgeProxy (\figurename~\ref{fig:dyn-node-ep}) successfully detected this change in processing latency (by measuring the actual latency) and acted promptly to remove \(s_3\) from the QoS pool so that requests are sent only to the remaining instances, i.e., \(s_1\) and \(s_4\). Also, it can be observed that in the time range T4-T5, when there was an increase in the network latency towards instance \(s_1\), proxy-mity still sent some of its requests to that instance due to the fact that it periodically refreshed the network latency information, while QEdgeProxy detected the increase in latency with the first requests that failed to meet the QoS requirements. It is important to emphasize that the time to adapt depends on the refresh frequency set in the proxy-mity configuration and the rate of the incoming requests, which indicates that this offset would be lower or non-existent when the request rate is higher than the refresh frequency.

\subsubsection{Computational and monitoring overhead}

We captured multiple snapshots of QEdgeProxy's resource consumption under high request load (\(\approx1000\) requests per second), revealing an average memory consumption of \textbf{10 MB}. Considering the K3s agent's memory footprint which ranges from 40 MB, as documented in an idle state on the Raspberry Pi 4 in \cite{cilicPerf}, to 80 MB that we measured in our experiment, we assert that the observed increase in usage remains within an acceptable range.
A higher resource consumption compared to proxy-mity which utilizes around 3 MB of memory is expected, as QEdgeProxy includes the features of maintaining the information on available service pods, QoS pools, and past QoS measurements.
\section{Conclusion and future work} 
\label{conclusion}

Extensive research has tackled service placement, replication, and deployment in the CC. However, ensuring uninterrupted data delivery from IoT devices to operating services in this dynamic environment remains a challenge. This paper introduces QEdgeProxy, an adaptive load balancing solution for routing client requests to IoT service instances in the CC, ensuring consistent QoS. By creating and managing a QoS pool of service instances, QEdgeProxy effectively balances load and continuously maintains QoS per individual clients. Our evaluation in a realistic K3s cluster shows QEdgeProxy to outperform both built-in Kubernetes load balancing and a state-of-the-art solution, serving a high rate of requests that meet the QoS requirements with minimal computational overhead. Future work will address challenges in large-scale deployments, including optimizing QoS pool maintenance and cluster monitoring, and exploring methods for neighboring proxies to share data for enhanced QoS prediction.

\section*{Acknowledgment}
This work has been supported by Croatian Science Foundation under the project IP-2019-04-1986 (IoT4us), and by the European Union’s Horizon program under the grant agreement No. 101079214 (AIoTwin).

\bibliographystyle{IEEEtran}
\bibliography{IEEEabrv,references.bib}

\begin{thebibliography}{10}
\providecommand{\url}[1]{#1}
\csname url@samestyle\endcsname
\providecommand{\newblock}{\relax}
\providecommand{\bibinfo}[2]{#2}
\providecommand{\BIBentrySTDinterwordspacing}{\spaceskip=0pt\relax}
\providecommand{\BIBentryALTinterwordstretchfactor}{4}
\providecommand{\BIBentryALTinterwordspacing}{\spaceskip=\fontdimen2\font plus
\BIBentryALTinterwordstretchfactor\fontdimen3\font minus \fontdimen4\font\relax}
\providecommand{\BIBforeignlanguage}[2]{{%
\expandafter\ifx\csname l@#1\endcsname\relax
\typeout{** WARNING: IEEEtran.bst: No hyphenation pattern has been}%
\typeout{** loaded for the language `#1'. Using the pattern for}%
\typeout{** the default language instead.}%
\else
\language=\csname l@#1\endcsname
\fi
#2}}
\providecommand{\BIBdecl}{\relax}
\BIBdecl

\bibitem{DBLP:conf/infocom/WojciechowskiOL21}
L.~Wojciechowski, K.~Opasiak, J.~Latusek, M.~Wereski, V.~Morales, T.~Kim, and M.~Hong, ``Netmarks: Network metrics-aware kubernetes scheduler powered by service mesh,'' in \emph{Proc. IEEE INFOCOM}, 2021.

\bibitem{DBLP:conf/netsoft/0001WVT19}
J.~Santos, T.~Wauters, B.~Volckaert, and F.~D. Turck, ``Towards network-aware resource provisioning in kubernetes for fog computing applications,'' in \emph{Proc. IEEE NetSoft}, 2019.

\bibitem{DBLP:journals/tiot/ToczeFPN23}
K.~Tocz{\'{e}}, A.~J. Fahs, G.~Pierre, and S.~Nadjm{-}Tehrani, ``Violinn: Proximity-aware edge placementwith dynamic and elastic resource provisioning,'' \emph{{ACM} Trans. Internet Things}, vol.~4, no.~1, 2023.

\bibitem{Sensors:Krivic}
P.~Krivic, M.~Kusek, I.~Čavrak, and P.~Skocir, ``Dynamic scheduling of contextually categorised internet of things services in fog computing environment,'' \emph{Sensors}, vol.~22, 01 2022.

\bibitem{DBLP:conf/ucc/PusztaiNMCRDVXZ22}
T.~W. Pusztai, S.~Nastic, A.~Morichetta, V.~Casamayor{-}Pujol, P.~Raith, S.~Dustdar, D.~Vij, Y.~Xiong, and Z.~Zhang, ``Polaris scheduler: {SLO-} and topology-aware microservices scheduling at the edge,'' in \emph{Proc. 15th {IEEE/ACM} UCC}, 2022.

\bibitem{kubernetes}
{Cloud Native Computing Foundation}, ``Kubernetes,'' \url{https://kubernetes.io/}.

\bibitem{7912240}
A.~Kapsalis, P.~Kasnesis, I.~S. Venieris, D.~I. Kaklamani, and C.~Z. Patrikakis, ``A cooperative fog approach for effective workload balancing,'' \emph{IEEE Cloud Computing}, vol.~4, no.~2, 2017.

\bibitem{DBLP:conf/ccgrid/FahsP19}
A.~J. Fahs and G.~Pierre, ``Proximity-aware traffic routing in distributed fog computing platforms,'' in \emph{Proc. IEEE/ACM CCGrid}, 2019.

\bibitem{s22082869}
Q.-M. Nguyen, L.-A. Phan, and T.~Kim, ``Load-balancing of kubernetes-based edge computing infrastructure using resource adaptive proxy,'' \emph{Sensors}, vol.~22, no.~8, 2022.

\bibitem{k3s}
{Cloud Native Computing Foundation}, ``K3s - lightweight kubernetes,'' \url{https://docs.k3s.io/}.

\bibitem{Rejiba21}
Z.~Rejiba, X.~Masip{-}Bruin, and E.~Mar{\'{\i}}n{-}Tordera, ``Towards user-centric, switching cost-aware fog node selection strategies,'' \emph{Future Gener. Comput. Syst.}, vol. 117, 2021.

\bibitem{Karagiannis23}
V.~Karagiannis, P.~A. Frangoudis, S.~Dustdar, and S.~Schulte, ``Context-aware routing in fog computing systems,'' \emph{{IEEE} Trans. Cloud Comput.}, vol.~11, no.~1, 2023.

\bibitem{Boban18}
M.~Boban, A.~Kousaridas, K.~Manolakis, J.~Eichinger, and W.~Xu, ``Connected roads of the future: Use cases, requirements, and design considerations for vehicle-to-everything communications,'' \emph{{IEEE} Veh. Technol. Mag.}, vol.~13, no.~3, 2018.

\bibitem{Alam23}
F.~Alam, A.~N. Toosi, M.~A. Cheema, C.~Cicconetti, P.~Serrano, A.~Iosup, Z.~Tari, and M.~Sarvi, ``Serverless vehicular edge computing for the internet of vehicles,'' \emph{{IEEE} Internet Comput.}, vol.~27, no.~4, 2023.

\bibitem{cilicPerf}
I.~Čilić, P.~Krivić, I.~Podnar~Žarko, and M.~Kušek, ``Performance evaluation of container orchestration tools in edge computing environments,'' \emph{Sensors}, vol.~23, no.~8, 2023.

\bibitem{network-policy}
``Kubernetes documentation: Network policies,'' https://kubernetes.io/docs/concepts/services-networking/network-policies/.

\bibitem{istio}
``Istio service mesh,'' https://istio.io/latest/about/service-mesh/.

\bibitem{imunes}
``{IMUNES:} integrated multiprotocol network emulator/simulator,'' http://imunes.net/.

\end{thebibliography}

\end{document}